\makeatletter
\def\input@path{{styles/nvidia/}}
\makeatother
\documentclass[10pt,logo,copyright]{nvidiatechreport}
\graphicspath{{styles/nvidia/}{figures/}}
\usepackage[T1]{fontenc}
\usepackage{amsmath,amssymb}
\usepackage{booktabs}
\usepackage{tabularx}
\usepackage{longtable}
\usepackage{float}
\usepackage{graphicx}
\usepackage{microtype}
\usepackage{natbib}
\usepackage{xcolor}
\usepackage{xspace}
\usepackage{xurl}
\usepackage{hyperref}

\definecolor{fpblue}{HTML}{1F77B4}

\newcommand{\method}{NVFP4-MLA\xspace}

\hypersetup{
  colorlinks=true,
  linkcolor=fpblue,
  citecolor=fpblue,
  urlcolor=fpblue
}

\newcommand{\papertitle}{NVFP4-MLA: Unified FP4 KV-Cache Storage and Attention Computation for \mbox{Multi-Head Latent Attention}}

\renewcommand{\today}{September 2026}

\title{\papertitle}
\author{Chao Ni}
\author{Qi Zhang}
\author{Xiao Wang}
\author{Zongfei Jing}
\author{Mingyang Hao}
\author{Binghan Chen}
\author{Kefeng Duan}
\author{Yu-Jung Chen}
\author{Mengdi Wang}
\author{Tijmen Blankevoort}
\author{Jun Yang}
\affil{NVIDIA}
\date{}

\begin{document}
\begin{abstract}
Long-context reasoning and agentic workloads make key-value (KV) cache capacity and attention computation the dominant costs of serving frontier large language models.
We present \method, the first design to push both KV-cache storage and attention computation of multi-head latent attention (MLA) to FP4.
\method co-designs the quantization algorithm with the attention kernel and the cache system.
Selective residual compensation preserves model accuracy by adding NVFP4 residuals only to the quantization-sensitive channels.
Two-dimensional quantization blocks make the shared latent cache compatible with the different reduction directions of the two attention batched matrix multiplications (BMMs), removing requantization between them.
On DeepSeek-R1 and Kimi-K3, \method matches BF16 MLA average accuracy across nine benchmarks while compressing the MLA KV cache by 3.20$\times$ and 3.05$\times$ relative to BF16 MLA.
On NVIDIA Rubin GPUs, \method speeds up the attention kernel over FP8 MLA by up to 2.82$\times$ on DeepSeek-R1 and 2.71$\times$ on Kimi-K3, and lowers p90 time to first token in agentic multi-turn serving by up to 3.5$\times$ and 2.1$\times$, respectively.
In disaggregated DeepSeek-R1 serving, it also raises per-GPU throughput by up to 2.30$\times$.
\end{abstract}

\maketitle
\section{Introduction}
\label{sec:introduction}

Multi-head latent attention (MLA) represents keys and values in a shared low-rank latent space and is the common attention pattern of several popular open-source large language models, including DeepSeek-V3~\citep{deepseekv3}, DeepSeek-R1~\citep{deepseekr1}, Kimi-K2.6, Kimi-K3~\citep{kimi2026k3}, and LongCat-Flash~\citep{longcat2025flash}, which use MLA in all or part of their attention layers.
The low-rank latent shrinks the key-value (KV) cache, but the cache still grows with sequence length and serving concurrency, and every decoding step reads all of it, which puts pressure on both memory capacity and bandwidth.
Agentic workloads sustain this pressure across turns and tool calls because they preserve and reuse growing contexts, which keeps the KV cache resident for longer.
When GPU memory cannot hold these caches, the server evicts them and recomputes the prefix in the next turn, which increases time to first token.
A smaller cache allows longer contexts, keeps more tokens and requests in GPU memory, reduces cache eviction and recomputation, and increases serving concurrency.

Beyond memory, computation can also be under pressure.
In grouped-query attention (GQA), each KV head serves only the few query heads in its group, whereas all MLA heads share the latent cache during decoding, so MLA performs more computation per cached byte (Appendix~\ref{app:intensity}).
Model-native multi-token prediction (MTP) verifies several queries against the same latent cache.
Reusing the cache across these queries raises arithmetic intensity further and makes MLA decoding compute-bound (Section~\ref{sec:eval-kernel}), so math throughput matters as much as cache bytes.
Together, these pressures motivate us to apply NVFP4, a 4-bit format that recent NVIDIA Tensor Cores support natively, to both KV-cache storage and attention computation in MLA, so that MLA benefits from smaller storage and higher math throughput.

\begin{figure}[t]
  \centering
  \includegraphics[width=\linewidth]{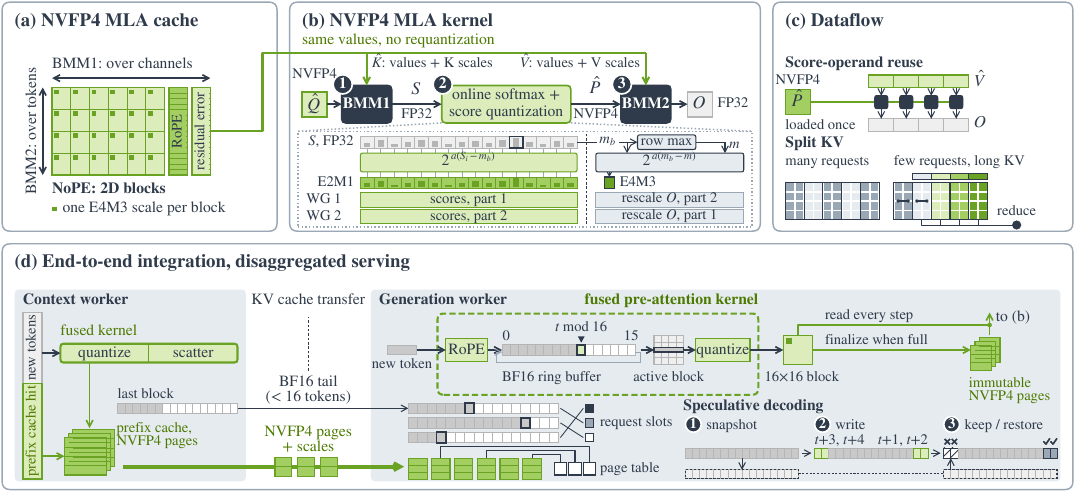}
  \caption{Overview of \method. (a) NVFP4 MLA cache: two-dimensional NoPE blocks shared by both BMMs, and RoPE with an NVFP4 residual. (b) NVFP4 MLA kernel: BMM1, online softmax fused with score quantization, and BMM2, with score construction and accumulator rescaling split across warp groups (WG). (c) Dataflow: score-operand reuse and split KV. (d) End-to-end integration in disaggregated serving: prefill, KV cache transfer, decode, and speculative decoding.}
  \label{fig:overview}
\end{figure}

We introduce \method, which brings unified NVFP4 storage and attention computation to MLA.
Its design addresses the challenges of applying FP4 quantization directly to MLA.
First, direct FP4 quantization degrades accuracy.
Outliers in the MLA cache concentrate in the last few channels, so selective residual compensation adds an NVFP4 residual only to these channels and keeps all elements in FP4 (Section~\ref{sec:eval-error-analysis}).
Second, MLA shares one latent representation between keys and values.
Default block-wise quantization groups this latent along channels, which matches the channel reduction of the first batched matrix multiplication (BMM1) but not the token reduction of BMM2.
\method introduces two-dimensional quantization blocks that are compatible with both reduction dimensions and therefore remove the need to requantize the shared latent cache for the second BMM (Section~\ref{sec:method-layout}).
Third, the higher arithmetic intensity makes math throughput a bottleneck.
\method therefore also quantizes queries and softmax scores online, so both BMMs run on native FP4 Tensor Cores and exploit the higher FP4 math throughput.
A corresponding cache system and kernel pipeline implement this design, and Figure~\ref{fig:overview} gives an overview.

\begin{samepage}
Our main contributions are:
\begin{enumerate}
  \item We design selective residual compensation, which keeps model accuracy by adding NVFP4 residuals to quantization-sensitive channels without mixed precision.
  \item We design two-dimensional quantization blocks and a corresponding cache system that let the shared quantized representation serve both BMMs without requantization.
  \item We build an efficient kernel pipeline. The pipeline quantizes queries and softmax scores online, so both BMMs benefit from the higher throughput of FP4 Tensor Cores.
\end{enumerate}
\end{samepage}

We evaluate \method on two frontier open MLA models, DeepSeek-R1 and Kimi-K3, across nine benchmarks that span reasoning, coding, instruction following, long context, and agentic tool use. \method matches BF16 MLA average accuracy on both models while reducing the MLA KV cache by 3.20$\times$ and 3.05$\times$ relative to BF16 MLA. On NVIDIA Rubin GPUs, \method speeds up the attention kernel over FP8 MLA by up to 2.82$\times$ on DeepSeek-R1 and 2.71$\times$ on Kimi-K3. In the AgentX benchmark, the smaller cache lowers p90 time to first token by up to 3.5$\times$ and 2.1$\times$, respectively. In disaggregated DeepSeek-R1 serving on the MLPerf and 8K/16K Reasoning workloads, per-GPU throughput reaches up to 2.30$\times$ that of FP8 MLA, and is 26.7\% and 39.9\% higher on average, respectively.

Section~\ref{sec:background} reviews MLA and the NVFP4 format.
Section~\ref{sec:method} presents selective residual compensation, two-dimensional quantization blocks, and the corresponding cache system and kernel pipeline.
Section~\ref{sec:evaluation} evaluates the accuracy, memory footprint, and performance of \method on DeepSeek-R1 and Kimi-K3.
Section~\ref{sec:related-work} compares \method with prior KV-cache quantization methods and quantized attention kernels.
Section~\ref{sec:conclusion} summarizes the findings.

\section{Background}
\label{sec:background}

\subsection{Multi-head latent attention}

\subsubsection{Low-rank compression creates shared KV state}

MLA stores one compressed latent vector for each token~\citep{deepseekv2,deepseekv3}. For a hidden state~$h_t$, a down-projection produces
\begin{equation}
  c_t = W^{\mathrm{DKV}}h_t,
  \qquad c_t\in\mathbb{R}^{d_c}.
  \label{eq:mla-latent}
\end{equation}
Learned up-projections recover head-specific keys and values from $c_t$. During inference, matrix absorption folds these projections into the query and output projections. The runtime therefore caches $c_t$ instead of materializing a separate key and value for every head.

The shared latent is involved in both BMMs. The first, BMM1 (QK), reduces over latent channels to generate attention logits, and the second, BMM2 (PV), reduces over cached tokens to generate the output. The same cached values are therefore reduced along two directions, so their block scales must serve both BMMs.

As Appendix~\ref{app:intensity} shows, MLA performs more computation per cached byte than a grouped-query attention (GQA) reference. Matrix absorption reduces cache traffic and reuses the latent cache for every query head. Model-native MTP raises arithmetic intensity further when the kernel reuses the cache across several query positions. MLA decoding therefore tends to be compute-bound.

\subsubsection{NoPE and RoPE channels serve different roles}

Applying rotary position encoding (RoPE)~\citep{su2021roformer} directly to the compressed keys would break matrix absorption, because the rotation depends on each token's position~\citep{deepseekv2}.
MLA therefore carries RoPE on a decoupled path, with additional per-head queries and a single key shared by all heads.
Each cached token thus holds a $d_c=512$-channel compressed latent, which we call NoPE, and a $d_{\mathrm R}=64$-channel decoupled key, which we call RoPE.
In BMM1, each head adds the NoPE and RoPE dot products to form its attention logit, whereas BMM2 aggregates only the NoPE latent.
Kimi-K3 keeps this decoupled layout but applies no positional encoding to either path~\citep{kimi2026k3}; we still call its decoupled channels RoPE.

\subsection{NVFP4 format and KV-cache quantization}

KV-cache quantization reduces per-token storage by encoding cached values with fewer bits~\citep{liu2024kivi}. To keep accuracy, prior methods redistribute large outliers through rotations, preserve selected outliers separately, or assign different bit widths across tensors and layers~\citep{hooper2024kvquant,li2025kvtuner,su2025rotatekv,zandieh2025turboquant}. Beyond accuracy, using the quantized cache directly in low-precision BMMs requires a compatible hardware format. We therefore use block-scaled NVFP4 for both cache storage and attention computation.

NVFP4 and MXFP4 both encode values in 4-bit E2M1 and attach a shared scale to each block~\citep{rouhani2023microscaling,nvidia2025nvfp4}. NVFP4 uses an E4M3 scale per 16 values, whereas MXFP4 uses an E8M0 power-of-two scale per 32 values. NVFP4's smaller blocks and non-power-of-two scales give finer control over local value ranges.

These formats have mostly been studied for weight and activation quantization~\citep{egiazarian2025microscaling}. For KV caches, their use also depends on how attention consumes the quantized values. When compute precision differs from storage precision, attention needs an extra dequantization step. Computing directly on the stored values avoids this step, but each BMM then requires the quantized values and scales in a compatible layout. This requirement links the block layout to the reduction axes of the attention BMMs. In GQA, separate key and value caches can adopt independent block layouts. In MLA, by contrast, both BMMs reduce the same latent cache along different axes, which couples their block layouts. A shared MLA cache intended for native FP4 computation therefore requires quantization blocks that support both reduction directions.

\section{Methodology}
\label{sec:method}

\method makes the shared latent a direct operand of both FP4 attention BMMs. Selective residual compensation reduces quantization error without mixed precision, two-dimensional quantization blocks align quantization with both reduction axes, the efficient kernel pipeline handles online quantization, and the corresponding cache system supports end-to-end integration.

\subsection{Quantization of Q, K, P, and V}
\label{sec:method-quantization}

During decode in quantized layers, \method quantizes all four attention operands: queries $Q$, keys $K$, softmax scores $P$, and values $V$. $P$ holds stable, unnormalized softmax scores, which \method quantizes online for BMM2.

$\mathcal{Q}_{1\times16}$ denotes NVFP4 quantization with contiguous 16-element blocks along the reduction axis of the consuming BMM: latent channels for queries and the RoPE cache, and cached tokens for softmax scores. $\mathcal{Q}_{16\times16}$ uses $16\times16$ token-by-channel blocks, each with a single scale (Appendix~\ref{app:nvfp4}).

For one query head and $T$ cached tokens, let $C=[C_{\mathrm{N}},C_{\mathrm{R}}]\in\mathbb{R}^{T\times(d_c+d_{\mathrm R})}$ and $q=[q_{\mathrm{N}},q_{\mathrm{R}}]\in\mathbb{R}^{1\times(d_c+d_{\mathrm R})}$. Let $S,\tilde{p}\in\mathbb{R}^{1\times T}$ denote attention logits and stable, unnormalized softmax scores, respectively; $\tilde{p}$ is the single-head instance of $P$. The normalized output is $O\in\mathbb{R}^{1\times d_c}$, and $d_{\mathrm{QK}}$ is the pre-absorption QK head width. NoPE uses $\mathcal{Q}_{16\times16}$ because both BMMs consume it, whereas RoPE enters only BMM1 and uses $\mathcal{Q}_{1\times16}$ along its channels. The quantized attention dataflow, evaluated as online softmax over successive KV tiles $\mathcal{T}_k$, is
\begin{equation}
  \begin{aligned}
    \widehat{C}_{\mathrm N} &= \mathcal{Q}_{16\times16}(C_{\mathrm N}),
    & \widehat{C}_{\mathrm R} &= \mathcal{Q}_{1\times16}(C_{\mathrm R}), \\
    \widehat{q}_{\mathrm N} &= \mathcal{Q}_{1\times16}(q_{\mathrm N}),
    & \widehat{q}_{\mathrm R} &= \mathcal{Q}_{1\times16}(q_{\mathrm R}), \\
    S &= \widehat{q}_{\mathrm N}\widehat{C}_{\mathrm N}^{\mathsf T}+\widehat{q}_{\mathrm R}\widehat{C}_{\mathrm R}^{\mathsf T},
    & m^{(k)} &= \max\bigl(m^{(k-1)},\,\max\nolimits_{i\in\mathcal{T}_k}S_i\bigr), \\
    \tilde{p}_i &= \exp\!\left((S_i-m^{(k)})/\sqrt{d_{\mathrm{QK}}}\right),
    & \widehat{p} &= \mathcal{Q}_{1\times16}(\tilde{p}), \\
    \ell^{(k)} &= \alpha_k\,\ell^{(k-1)}+\sum_{i\in\mathcal{T}_k}\tilde{p}_i,
    & \widetilde{O}^{(k)} &= \alpha_k\,\widetilde{O}^{(k-1)}+\sum_{i\in\mathcal{T}_k}\widehat{p}_i\,\widehat{C}_{\mathrm N,i}, \\
    \alpha_k &= \exp\!\left((m^{(k-1)}-m^{(k)})/\sqrt{d_{\mathrm{QK}}}\right),
    & O &= \widetilde{O}/\ell.
  \end{aligned}
  \label{eq:base}
\end{equation}
Here $\mathcal{T}_k$ is the $k$-th KV tile and $i\in\mathcal{T}_k$, with $m^{(0)}=-\infty$ and $\ell^{(0)}=\widetilde{O}^{(0)}=0$; $m$, $\ell$, and $\widetilde{O}$ without a superscript are the running values, and $O$ is taken after the last tile. Both BMMs accumulate in FP32, and softmax keeps $S$, $m$, $\ell$, and the final division by $\ell$ in FP32.

\subsection{Selective residual compensation}
\label{sec:method-residual}

FP4 quantization error in the MLA cache is disproportionately large in the RoPE partition (Section~\ref{sec:eval-error-analysis}). \method therefore keeps a single NVFP4 component for NoPE but adds a second NVFP4 component only for the RoPE quantization error.
Because the RoPE partition holds 64 of the 576 latent channels, the residual adds little storage and computation; Section~\ref{sec:eval-accuracy} evaluates the accuracy of \method.

The native kernel packs RoPE and its quantization error separately with independent NVFP4 block scales, logically concatenates them along the reduction dimension of BMM1, and duplicates the corresponding channels. The compensated logits, which replace $S$ in Equation~\ref{eq:base}, therefore take a single BMM1 reduction, without merging base and residual into a high-precision tensor.

\subsection{Two-dimensional quantization blocks for shared keys and values}
\label{sec:method-layout}

\begin{figure}[tb]
  \centering
  \includegraphics[width=0.51\linewidth]{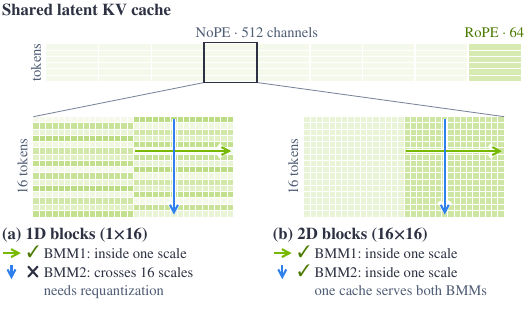}
  \caption{Two-dimensional quantization blocks. Both BMMs consume the same FP4 values and scales without requantization. One-dimensional $1\times16$ blocks (a) serve only BMM1, whereas $16\times16$ blocks (b) serve both. \method uses (b) for NoPE and (a) for RoPE, which only BMM1 consumes.}
  \label{fig:two-dimensional-blocks}
\end{figure}

As Section~\ref{sec:background} shows, the shared MLA latent requires scales that work in both reduction directions. A conventional one-dimensional layout aligns its blocks with BMM1 but not with native NVFP4 BMM2 (Figure~\ref{fig:two-dimensional-blocks}a). Converting this layout requires dequantizing the cached E2M1 values, transposing them, and requantizing them into BMM2-aligned blocks. Requantization introduces additional quantization error. It also adds overhead such as memory traffic, a scale reduction, and synchronization before BMM2.

\method therefore quantizes the shared NoPE latent in two-dimensional $16\times16$ blocks over tokens and channels (Figure~\ref{fig:two-dimensional-blocks}b). One scale covers every 16-element row and column segment of a block, so the block is valid in either reduction direction: both BMMs share the same NVFP4 values and scales without a high-precision intermediate or a requantization pass.

\subsection{Kernel implementation}
\label{sec:implementation}

To turn FP4's higher Tensor Core throughput into a faster attention kernel, \method quantizes queries and softmax scores online (Figure~\ref{fig:overview}b) and applies four optimizations: fusing online softmax with FP4 score quantization, balancing FP4 score construction and accumulator rescaling, reusing FP4 score operands across BMM2 output tiles, and shape-adaptive thread-block scheduling.

\textbf{Fusing online softmax with FP4 score quantization.}
Online FP4 score quantization makes the logit-to-score path heavier than in FP8 MLA because the path must also derive block scales and pack the FP4 operand for BMM2. We algebraically fuse online softmax, FP4 score quantization, and scale generation. As in online softmax, let $m$ be the running maximum of one row of logits $S$, and let $a=\log_2(e)/\sqrt{d_{\mathrm{QK}}}$, so that $\tilde{p}_i=2^{a(S_i-m)}$ in Equation~\ref{eq:base}. Let $m_b=\max_{j\in b}S_j\le m$; for each 16-element block $b$ and $i\in b$,
\begin{equation}
  \begin{aligned}
    \tilde{p}_i &= 2^{a(m_b-m)}\,2^{a(S_i-m_b)},
    \qquad \widehat{p}_i=\gamma_P\,s_b\,x_i, \\
    x_i &= \mathcal{Q}_{\mathrm{E2M1}}\bigl(M_{\mathrm{E2M1}}\,2^{a(S_i-m_b)}\bigr), \\
    s_b &= \mathcal{Q}_{\mathrm{E4M3}}\bigl(M_{\mathrm{E4M3}}\,2^{a(m_b-m)}\bigr),
  \end{aligned}
  \label{eq:p-fma-form}
\end{equation}
where $\mathcal{Q}_{\mathrm{E2M1}}$ and $\mathcal{Q}_{\mathrm{E4M3}}$ round to E2M1 and E4M3, $M_{\mathrm{E2M1}}=6$ and $M_{\mathrm{E4M3}}=448$ are the largest E2M1 and finite E4M3 magnitudes (Appendix~\ref{app:nvfp4}), $\gamma_P=1/(M_{\mathrm{E2M1}}M_{\mathrm{E4M3}})$ is a constant tensor scale, and $x_i$ is quantized against the unrounded block scale. Because $S_i\le m_b\le m$, the two arguments lie in $(0,M_{\mathrm{E2M1}}]$ and $(0,M_{\mathrm{E4M3}}]$, so neither rounding saturates. The E2M1 value $x_i$ depends on $S_i$ and $m_b$ but not on $m$, so the kernel evaluates the exponent with a fused multiply-add, converts the result to E2M1, and packs it before the cross-block reduction produces $m$. Once $m$ is available, only the block scale $s_b$ remains (Appendix~\ref{app:probability-scaling}).

\textbf{Balancing FP4 score construction and accumulator rescaling.}
Fusion shortens the logit-to-score path, but this stage still carries more non-matrix work than its FP8 MLA counterpart. Rather than dedicate separate warp groups to FP4 score construction and online-softmax accumulator rescaling, we partition each logit tile across the warp groups. Each group constructs the FP4 score operand for its logit partition and also rescales a complementary partition of the resident BMM2 accumulator when the running maximum changes. This mixed assignment keeps the logit-to-score stage from serializing the pipeline.

\textbf{Reusing FP4 score operands across BMM2 output tiles.}
Each packed FP4 score slice contributes to multiple BMM2 output tiles. The kernel uses the Tensor Core operand-reuse mechanism to retain this $\widehat{P}$ operand of BMM2 across these tiles instead of reloading it for each tile, so more execution bandwidth remains for FP4 score construction.

\textbf{Shape-adaptive CTA scheduling.}
With MTP, each request has more queries, so the kernel can use larger matrix multiply-accumulate (MMA) tiles. It gives each cluster of thread blocks (cooperative thread arrays, CTAs) the packed query rows of one request; the CTAs of a cluster split the KV tiles, exchange the FP4 scores through distributed shared memory, and use the same FP4 operands and softmax. At long KV lengths, the kernel also splits the KV of a request into pieces that run on idle streaming multiprocessors (Figure~\ref{fig:overview}c) and merges them with a reduction kernel.

\subsection{End-to-end integration}
\label{sec:integration}

For end-to-end integration, we adapt the cache system to be compatible with two-dimensional quantization blocks, the NVFP4 attention kernel, paged KV caches, and speculative decoding (Figure~\ref{fig:overview}d).
During prefill, a fused quantize-and-scatter kernel writes completed NVFP4 blocks into paged storage, which decode reads without requantization.
Disaggregated serving transfers only these pages, namely the E2M1 values and their scales, plus the BF16 tail of the incomplete last block.
During decode, a fused pre-attention kernel applies RoPE where the model uses it, updates a 16-token BF16 ring, quantizes the operands, and writes the cache. Each request owns a stable slot in the ring, and the runtime refreshes the active NVFP4 block from BF16 values and finalizes the block once it fills. It therefore never has to dequantize a block, append newly generated tokens, and requantize it, which would introduce additional quantization error.
For speculative decoding such as MTP, the runtime snapshots the BF16 ring before speculative writes, commits accepted candidates, and restores locations overwritten by rejected candidates.
Completed NVFP4 pages remain unchanged during rollback, so rejected tokens cannot contaminate persistent state.
\method has been integrated into TensorRT-LLM~\citep{nvidia2026tensorrtllm}.

\section{Evaluation}
\label{sec:evaluation}

We evaluate \method on DeepSeek-R1 and Kimi-K3 in three parts: accuracy (Section~\ref{sec:eval-accuracy}), KV-cache memory reduction (Section~\ref{sec:eval-storage}), and performance (Section~\ref{sec:eval-performance}).

\subsection{Experimental setup}
\label{sec:eval-setup}

\textbf{Accuracy.}
Both models use the same nine tasks, spanning reasoning, coding, instruction following, long context, and agentic tool use (Table~\ref{tab:accuracy}).
RULER and MRCR use a 128K-token context.
We compare \method with BF16 MLA, FP8 MLA, and vanilla NVFP4 MLA.
\textbf{BF16 MLA} keeps the KV cache and attention BMM computation in BF16.
\textbf{FP8 MLA} quantizes the KV cache and attention BMM computation to E4M3.
\textbf{Vanilla NVFP4 MLA} quantizes them directly to NVFP4 without residual compensation.

\textbf{Storage.}
We count the stored MLA cache. DeepSeek-R1 uses NVFP4 in all 61 MLA layers. Kimi-K3 uses NVFP4 in 22 of its 24 MLA layers and FP8 in the final two; the count excludes the recurrent state of its 69 Kimi Delta Attention (KDA) layers.

\textbf{Performance.}
All performance experiments run on NVIDIA Rubin GPUs, and serving runs use TensorRT-LLM with the same runtime and the same set of parallel layouts for both precisions.
On DeepSeek-R1 with MTP=3, MLPerf uses the evaluation prompts of the MLPerf Inference~\citep{reddi2020mlperf} DeepSeek-R1 benchmark, and 8K/16K Reasoning uses requests of about 8K input and 16K output tokens. Both workloads use disaggregated serving, with generation under attention data and expert parallelism (DEP) or attention tensor and expert parallelism (TEP).
AgentX replays 393 recorded multi-turn Claude Code sessions from the SemiAnalysis InferenceX benchmark on both models, using distributed weight data parallelism (DWDP)~\citep{nvidia2026dwdp} for Kimi-K3.
Attention kernel measurements cover 39 shapes per model, at KV lengths of 2K--128K and per-GPU batches of 8--1,024, with MTP=3 on DeepSeek-R1 and MTP=7 on Kimi-K3.

\subsection{Accuracy}
\label{sec:eval-accuracy}

\method matches BF16 MLA average accuracy on both models (Section~\ref{sec:eval-downstream}), and white-box analysis of the cache shows that direct FP4 quantization loses precision on significant outliers in the RoPE channels, which the NVFP4 residual encodes more accurately than FP8 (Section~\ref{sec:eval-error-analysis}).

\subsubsection{Downstream accuracy}
\label{sec:eval-downstream}

With NVFP4 attention BMM computation, \method matches BF16 MLA average accuracy on both models (59.82 vs.\ 59.80 and 80.28 vs.\ 80.27; Table~\ref{tab:accuracy}), compared with $-0.06$ and $-0.33$ percentage points (pp) for FP8 MLA.

\begin{table}[!tbp]
  \caption{Downstream accuracy on DeepSeek-R1 and Kimi-K3. \method matches BF16 MLA average accuracy and greatly reduces the accuracy drop of vanilla NVFP4 MLA. Scores in \%; Avg: nine-task mean. Repeats per task: 64 for GPQA~\citep{rein2024gpqa}, AIME 2025, and AA-LCR; 8 for LiveCodeBench~\citep{jain2025livecodebench}, SciCode~\citep{tian2024scicode}, and $\tau^2$-Bench~\citep{barres2026tau2bench}; 5 for IFBench~\citep{pyatkin2025ifbench}; 3 for MRCR~\citep{vodrahalli2024michelangelo}; 1 for RULER~\citep{hsieh2024ruler}.}
  \label{tab:accuracy}
  \centering
  \fontsize{8}{9}\selectfont
  \renewcommand{\arraystretch}{1.0}
  \setlength{\tabcolsep}{1.2pt}
  \begin{tabular}{@{}>{\raggedright\arraybackslash}p{20mm}*{10}{>{\centering\arraybackslash}p{\dimexpr(\linewidth-20mm-20\tabcolsep)/10\relax}}@{}}
  \toprule
  Configuration & GPQA & IFBench & \shortstack{AIME\\2025} & \shortstack{LiveCode\\Bench v6} & SciCode & RULER & MRCR & AA-LCR & $\tau^2$-Bench & Avg \\
  \midrule
  \multicolumn{11}{@{}l}{\textbf{DeepSeek-R1}} \\
  \addlinespace[2pt]
  BF16 MLA & 80.07 & 39.53 & 87.14 & 71.31 & 42.60 & 92.97 & 36.35 & 52.52 & 35.75 & 59.80 \\
  \midrule[\lightrulewidth]
  FP8 MLA & 80.16 & 37.93 & 87.60 & 71.75 & 42.27 & 92.67 & 37.21 & 52.62 & 35.42 & 59.74 \\
  \midrule[\lightrulewidth]
  \shortstack[l]{Vanilla NVFP4\\MLA} & 77.38 & 34.87 & 73.80 & 62.11 & 40.64 & 90.52 & 29.54 & 46.06 & 31.91 & 54.09 \\[3pt]
  \method & 79.83 & 38.07 & 85.94 & 71.78 & 43.12 & 92.77 & 36.95 & 51.86 & 38.05 & 59.82 \\
  \midrule[\heavyrulewidth]
  \multicolumn{11}{@{}l}{\textbf{Kimi-K3}} \\
  \addlinespace[2pt]
  BF16 MLA & 93.66 & 73.96 & 96.20 & 57.49 & 58.48 & 97.00 & 85.02 & 81.75 & 78.84 & 80.27 \\
  \midrule[\lightrulewidth]
  FP8 MLA & 93.50 & 73.69 & 95.63 & 57.32 & 58.19 & 97.06 & 84.38 & 82.19 & 77.49 & 79.94 \\
  \midrule[\lightrulewidth]
  \shortstack[l]{Vanilla NVFP4\\MLA} & 88.26 & 61.79 & 87.71 & 10.68 & 2.27 & 95.98 & 75.21 & 77.31 & 74.89 & 63.79 \\[3pt]
  \method & 93.43 & 74.91 & 96.20 & 57.30 & 58.09 & 97.11 & 84.10 & 82.31 & 79.06 & 80.28 \\[3pt]
  \bottomrule
  \end{tabular}
\end{table}

\subsubsection{Quantization error analysis}
\label{sec:eval-error-analysis}

\textbf{Measurement.}
To locate FP4 error in the MLA cache, we analyze dumped BF16 cache tensors across layers of DeepSeek-R1 and MLA layers of Kimi-K3.
We measure reconstruction error as the normalized mean squared error $\mathrm{NMSE}=\lVert \widehat{X}-X\rVert_F^2/\lVert X\rVert_F^2$.
NoPE stays at NVFP4, and only the RoPE encoding varies: NVFP4, FP8 E4M3, or NVFP4 plus residual.

\begin{figure}[!htbp]
  \centering
  \includegraphics[width=\linewidth]{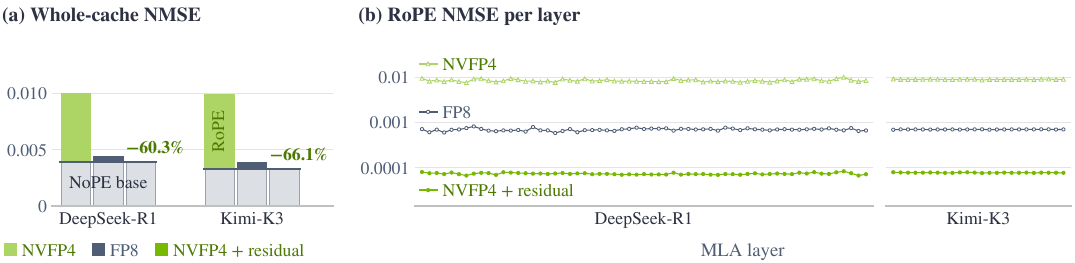}
  \caption{Cache quantization error. A RoPE-only NVFP4 residual lowers whole-cache NMSE and is more accurate than FP8 RoPE in every layer. NoPE uses NVFP4; RoPE uses NVFP4, FP8 E4M3, or NVFP4 plus residual. (a) Whole-cache NMSE averaged over MLA layers. (b) Per-layer NMSE of each RoPE encoding (log scale).}
  \label{fig:cache-error}
\end{figure}

\begin{figure}[!htbp]
  \centering
  \includegraphics[width=\linewidth]{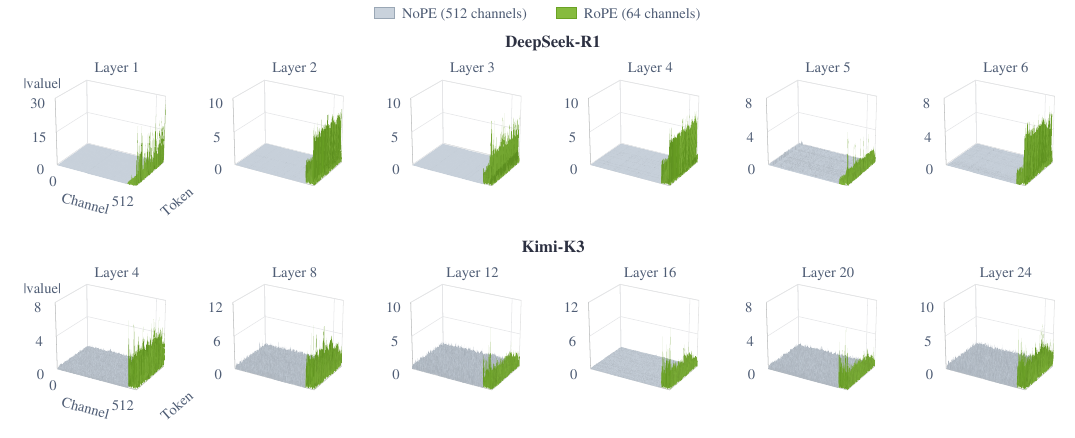}
  \caption{RoPE and NoPE magnitudes. RoPE channels peak at 3.9--125$\times$ the NoPE maximum.}
  \label{fig:cache-surfaces}
\end{figure}

\textbf{Error concentrates in RoPE.}
RoPE holds 64 of the 576 channels, yet compensating it alone lowers mean whole-cache NMSE by 60.3\% on DeepSeek-R1 and 66.1\% on Kimi-K3 (Figure~\ref{fig:cache-error}a).
Figure~\ref{fig:cache-surfaces} contrasts the two partitions in the first few MLA layers of both models: the RoPE channels peak at 3.9--125$\times$ the NoPE maximum and vary sharply across tokens, whereas the NoPE latent stays flat.

\textbf{NVFP4 residual versus FP8 RoPE.}
The NVFP4 residual encodes RoPE more accurately than FP8 E4M3. Relative to plain NVFP4, it reduces mean RoPE NMSE by factors of 115.4 on DeepSeek-R1 and 116.1 on Kimi-K3, whereas FP8 RoPE leaves 9.5$\times$ and 9.1$\times$ as much error as the residual on average and 7.9--11.2$\times$ and 8.7--9.3$\times$ as much in every layer (Figure~\ref{fig:cache-error}b).

\subsubsection*{Takeaway}
\method matches BF16 MLA average accuracy on both models while running both attention BMMs in NVFP4, whereas vanilla NVFP4 MLA drops by 5.71 and 16.48 pp. White-box analysis shows that this drop stems from significant outliers in the RoPE channels. Selective residual compensation therefore mitigates the accuracy drop with a small overhead (Section~\ref{sec:eval-storage}).

\subsection{KV-cache memory reduction}
\label{sec:eval-storage}

Including the residual, \method compresses the stored MLA cache by 3.20$\times$ and 3.05$\times$ relative to BF16 MLA and by 1.60$\times$ and 1.52$\times$ relative to FP8 MLA on DeepSeek-R1 and Kimi-K3 (Table~\ref{tab:storage}). The RoPE residual accounts for only 10\% of each NVFP4 layer's cache.

\begin{table}[!htbp]
  \caption{Stored MLA cache size. \method compresses the stored MLA cache by 3.20$\times$ and 3.05$\times$ relative to BF16 MLA and by 1.60$\times$ and 1.52$\times$ relative to FP8 MLA. Bytes per cached token, summed over all MLA layers.}
  \label{tab:storage}
  \centering
  \small
  \setlength{\tabcolsep}{7pt}
  \begin{tabularx}{\linewidth}{@{}>{\raggedright\arraybackslash}Xrrrr@{}}
    \toprule
    & \multicolumn{2}{c}{DeepSeek-R1} & \multicolumn{2}{c}{Kimi-K3} \\
    \cmidrule(lr){2-3}\cmidrule(l){4-5}
    Cache representation & Bytes/token & Reduction & Bytes/token & Reduction \\
    \midrule
    BF16 MLA & 70,272 & 1.00$\times$ & 27,648 & 1.00$\times$ \\
    FP8 MLA & 35,136 & 2.00$\times$ & 13,824 & 2.00$\times$ \\
    \method & 21,960 & 3.20$\times$ & 9,072 & 3.05$\times$ \\
    \bottomrule
  \end{tabularx}
\end{table}

On Kimi-K3, the MLA cache shares the KV budget with the recurrent state of the KDA layers; under the same budget, \method holds 47--49\% more cached tokens than FP8 MLA, depending on the share taken by KDA state. Section~\ref{sec:eval-agentx} shows how this capacity turns into prefix reuse.

\subsection{Performance}
\label{sec:eval-performance}

We evaluate performance at three levels. In agentic serving, the smaller cache of \method keeps more session histories and lowers p90 time to first token by up to 3.5$\times$ on DeepSeek-R1 and by 2.1$\times$ on Kimi-K3 (Section~\ref{sec:eval-agentx}). Kernel measurements show an attention kernel that runs up to 2.82$\times$ and 2.71$\times$ faster than FP8 MLA on DeepSeek-R1 and Kimi-K3 (Section~\ref{sec:eval-kernel}). In disaggregated DeepSeek-R1 serving, the faster kernel raises per-GPU throughput by up to 2.30$\times$ (Section~\ref{sec:eval-serving}).

\subsubsection{Prefix-cache reuse}
\label{sec:eval-agentx}

With its smaller cache (Section~\ref{sec:eval-storage}), \method delays cache eviction as the number of concurrent sessions grows. Agentic sessions benefit from this capacity because context workers retain earlier turns' KV state and prefill only newly appended tokens.

\textbf{DeepSeek-R1.}
We replay the AgentX sessions of Section~\ref{sec:eval-setup} under an equal KV-cache budget of 43\,GiB per context GPU, in which \method holds 60\% more cached tokens than FP8 MLA (Section~\ref{sec:eval-storage}) and the trace allows a hit rate of about 94--96\%.
FP8 MLA starts evicting at 32 sessions (Table~\ref{tab:agentx-serving}): from 32 to 64 sessions, its hit rate falls from 87.6\% to 54.1\% and its p90 time to first token (TTFT) rises from 11.9 to 178\,s.
\method keeps a higher hit rate and therefore repeats less prefill: with 48 sessions, it raises the hit rate from 77.0\% to 87.1\% and lowers p90 TTFT from 108 to 30.9\,s, a 3.5$\times$ reduction, and the reduction is 1.5--3.5$\times$ across all points where FP8 MLA evicts.

\begin{table}[!htbp]
  \caption{AgentX serving. At equal KV-cache memory, \method keeps higher cache-hit rates and lower p90 TTFT once FP8 MLA starts evicting histories. AgentX replay on Rubin GPUs. Kimi-K3 KV-cache memory also holds KDA state. Concurrent sessions: agent sessions kept alive; Cache hit: fraction of input tokens reused from the prefix cache.}
  \label{tab:agentx-serving}
  \centering
  \small
  \begin{tabular}{lrrrrr}
    \toprule
     & & \multicolumn{2}{c}{p90 TTFT (s)} & \multicolumn{2}{c}{Cache hit (\%)} \\
    \cmidrule(lr){3-4}\cmidrule(lr){5-6}
    Model & Concurrent sessions & FP8 MLA & \method & FP8 MLA & \method \\
    \midrule
    DeepSeek-R1 & 16 & 2.98 & 1.99 & 95.2 & 95.1 \\
                & 24 & 2.68 & 2.41 & 93.6 & 94.4 \\
                & 32 & 11.9 & 5.36 & 87.6 & 92.4 \\
                & 40 & 36.6 & 23.7 & 82.3 & 87.9 \\
                & 48 & 108 & 30.9 & 77.0 & 87.1 \\
                & 64 & 178 & 79.8 & 54.1 & 79.6 \\
    \midrule
    Kimi-K3     & 128 & 11.0 & 5.24 & 90.8 & 94.4 \\
    \bottomrule
  \end{tabular}
\end{table}

\textbf{Kimi-K3.}
We replay the same sessions on Kimi-K3 under an equal KV-cache budget of 47.4\,GiB per context GPU, in which \method holds 47--49\% more cached tokens than FP8 MLA (Section~\ref{sec:eval-storage}) and the trace allows a hit rate of at most 95.4\%.
FP8 MLA already evicts with 128 sessions (Table~\ref{tab:agentx-serving}).
\method keeps a higher hit rate and therefore repeats less prefill: it raises the hit rate from 90.8\% to 94.4\% and lowers p90 TTFT from 11.0 to 5.24\,s, a 2.1$\times$ reduction.

\subsubsection{Kernel performance}
\label{sec:eval-kernel}

Besides its smaller cache, \method shortens the attention computation itself, because both BMMs run on FP4 Tensor Cores. Its speedup over FP8 MLA grows with KV length, reaching 2.82$\times$ on DeepSeek-R1 and 2.71$\times$ on Kimi-K3.

\textbf{MTP and the roofline.}
MTP=3 reuses each cache read across four query positions (Appendix~\ref{app:intensity}), which raises the arithmetic intensity of FP8 MLA from 484 to 1,934 FLOPs per cache byte, beyond the 911\,FLOP/B ridge point of a Rubin GPU; FP8 MLA decoding with MTP is therefore compute-bound. A storage-only alternative that dequantizes an NVFP4 cache to FP8 operands performs the same FP8 MMAs and softmax as FP8 MLA and shortens only cache reads that no longer limit the kernel, so it cannot finish faster than FP8 MLA. \method is faster because its FP4 BMMs shorten the computation itself.

\begin{figure}[!tb]
  \centering
  \includegraphics[width=\linewidth]{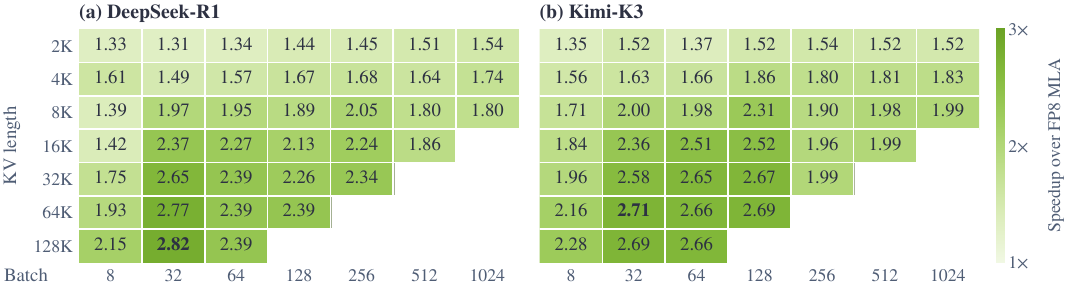}
  \caption{Kernel speedup over FP8 MLA on one Rubin GPU. (a) DeepSeek-R1. (b) Kimi-K3.}
  \label{fig:kernel-speedup}
\end{figure}

\textbf{KV-length and batch sweep.}
Each measurement profiles a single attention kernel under a CUDA graph, with equal KV lengths across requests, and includes the split-KV reduction. FP8 MLA uses its faster split-KV strategy per shape; \method uses its best CTA scheduling per shape.
\method is faster at every measured shape (Figure~\ref{fig:kernel-speedup}), by 1.31--2.82$\times$ with a geometric mean of 1.87$\times$ on the 39 DeepSeek-R1 shapes and by 1.35--2.71$\times$ with a geometric mean of 1.99$\times$ on the 39 Kimi-K3 shapes.
The speedup generally grows with KV length, from 1.3--1.5$\times$ at 2K, where fixed per-call cost dominates, to 2.2--2.8$\times$ at 128K, where per-token attention work dominates.

\subsubsection{End-to-end serving}
\label{sec:eval-serving}

\method improves the throughput--latency trade-off of disaggregated DeepSeek-R1 serving, with gains that grow with context length.

\textbf{Pareto measurement.}
Closed-loop clients sweep the number of in-flight requests for each generation layout, and each precision's Pareto frontier takes, at every per-user generation rate, the highest per-GPU throughput of any of its layouts.

\begin{figure}[!tb]
  \centering
  \includegraphics[width=\linewidth]{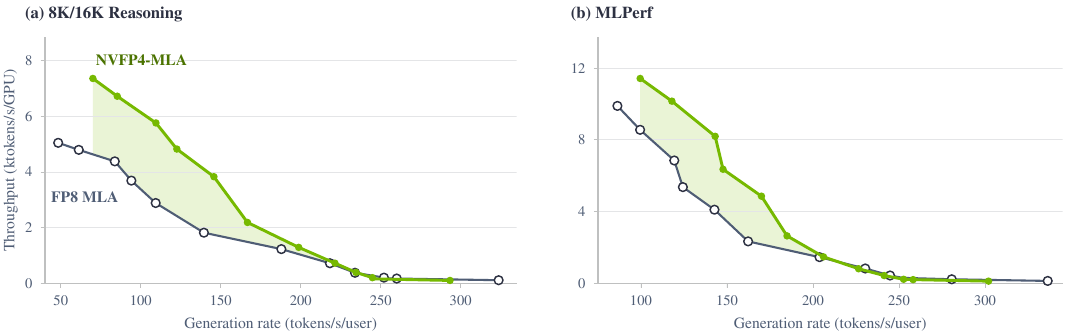}
  \caption{Serving Pareto frontiers. \method raises per-GPU throughput over most of the range. Pareto frontiers for DeepSeek-R1 disaggregated serving on Rubin GPUs with MTP=3: (a) 8K/16K Reasoning; (b) MLPerf.}
  \label{fig:pareto-measured}
\end{figure}

\textbf{Throughput.}
At the same per-user generation rate, \method serves more output tokens per GPU over most of the measured range (Figure~\ref{fig:pareto-measured}): 26.7\% more on MLPerf and 39.9\% more on 8K/16K Reasoning on average, and up to 2.30$\times$ and 2.25$\times$ as many.

\textbf{Parallel layouts.}
Both precisions reach their highest throughput with DEP16 up to about 250 tokens/s/user and with TEP8 at higher per-user rates.

\subsubsection*{Takeaway}
\method improves performance through two effects. Its smaller cache keeps more session histories, lowering p90 TTFT by up to 3.5$\times$ on DeepSeek-R1 and 2.1$\times$ on Kimi-K3 when GPU memory bounds the prefix cache. Its FP4 BMMs shorten attention computation, with speedups that grow with context length, reaching 2.82$\times$ and 2.71$\times$ at the kernel level and up to 2.30$\times$ in per-GPU throughput for DeepSeek-R1 serving.

\section{Related work}
\label{sec:related-work}

\textbf{KV-cache quantization.}
Most previous work focuses on quantizing the KV cache to reduce its footprint and memory traffic. KIVI~\citep{liu2024kivi} quantizes keys per channel and values per token, keeping the most recent tokens in full precision. KVQuant~\citep{hooper2024kvquant} quantizes keys per channel before RoPE and keeps outliers in a sparse format. KVTuner~\citep{li2025kvtuner} searches layer-wise mixed precision for keys and values. RotateKV~\citep{su2025rotatekv} and TurboQuant~\citep{zandieh2025turboquant} rotate cached vectors to spread outliers before low-bit quantization. Serving systems such as Atom~\citep{zhao2024atom} and QServe~\citep{lin2025qserve} also store 4-bit KV caches alongside low-bit weights and activations. These methods target separate key and value caches rather than MLA's shared latent representation. They also execute the attention BMMs at higher precision, which adds dequantization overhead for the cached values and limits MLA decoding once MTP makes it compute-bound (Section~\ref{sec:eval-kernel}). \method instead stores a shared MLA cache in NVFP4 whose scale blocks match both reduction directions, so both BMMs read it in FP4 without dequantization and run at FP4 Tensor Core throughput.

\textbf{Quantized attention kernels.}
Low-precision attention kernels quantize attention operands to run the attention BMMs on faster low-precision Tensor Cores. SageAttention~\citep{zhang2024sageattention} quantizes queries and keys to INT8 for the $QK^\top$ BMM and keeps the $PV$ BMM mostly in FP16. SageAttention3~\citep{zhang2025sageattention3} runs both BMMs in NVFP4, quantizing FP16 queries, keys, and values inside each attention call, and mainly targets image and video generation. SnapMLA~\citep{zhang2026snapmla} pushes MLA computation to FP8 with a per-token FP8 latent cache and keeps the RoPE channels in BF16. \method pushes it further to FP4 and combines low-precision storage and computation: both BMMs consume the persistent NVFP4 MLA cache without dequantization, and the RoPE channels are quantized to NVFP4 with residual compensation rather than kept in high precision.

\textbf{Evaluation scope.}
Prior KV-cache quantization studies evaluate non-MLA models of limited size, mostly with perplexity, short-generation tasks, or long-context retrieval~\citep{liu2024kivi,hooper2024kvquant,li2025kvtuner}. These settings do not test long reasoning traces or multi-turn agentic sessions, where quantization error can accumulate over thousands of generated tokens. We instead evaluate large MLA reasoning models with a broader set of tests, spanning downstream accuracy on reasoning and agentic tasks, cache footprint, and kernel and end-to-end serving performance, to understand how \method affects both model quality and efficiency.

\section{Conclusion}
\label{sec:conclusion}

We have presented \method, which shows that NVFP4 can serve as both the storage format and the compute format of MLA. Selective residual compensation targets quantization outliers and greatly reduces the accuracy drop of direct FP4 quantization, and two-dimensional quantization blocks let one shared latent cache supply both BMMs. We also design a corresponding cache system and an efficient kernel pipeline. \method matches BF16 MLA average accuracy on DeepSeek-R1 and Kimi-K3, speeds up the attention kernel over FP8 MLA by up to 2.82$\times$ and 2.71$\times$, lowers p90 time to first token in agentic multi-turn serving by up to 3.5$\times$ and 2.1$\times$, and raises per-GPU throughput in disaggregated DeepSeek-R1 serving by 26.7\% and 39.9\% on average on two workloads and by up to 2.30$\times$. By computing attention in FP4 while preserving accuracy, \method turns KV-cache compression into real inference speedups.

\clearpage
\appendix
\section{NVFP4 quantization}
\label{app:nvfp4}

Let $\mathbb{F}_{\mathrm{E2M1}}$ be the set of E2M1 values, with largest magnitude $M_{\mathrm{E2M1}}=6$, and let $\mathbb{F}_{\mathrm{E4M3}}$ be the set of finite E4M3 values, with largest magnitude $M_{\mathrm{E4M3}}=448$. Let $\mathcal{Q}_{\mathrm{E2M1}}$ and $\mathcal{Q}_{\mathrm{E4M3}}$ round a real number onto $\mathbb{F}_{\mathrm{E2M1}}$ and $\mathbb{F}_{\mathrm{E4M3}}$, respectively, with saturation; $\mathcal{Q}_{\mathrm{E2M1}}$ rounds to nearest with ties to even. An NVFP4 tensor stores an FP32 tensor scale $\gamma>0$, one E4M3 scale $s_b$ for each block $b$ of 16 elements, and one E2M1 value $x_i$ for each element. For a real tensor $X$ and a partition $\mathcal{B}$ of its indices into 16-element blocks, NVFP4 quantization $\mathcal{Q}(\,\cdot\,;\mathcal{B})$ is
\begin{equation}
  s_b=\mathcal{Q}_{\mathrm{E4M3}}\!\left(\frac{\max_{j\in b}\lvert X_j\rvert}{M_{\mathrm{E2M1}}\gamma}\right),\qquad
  x_i=\mathcal{Q}_{\mathrm{E2M1}}\!\left(\frac{X_i}{\gamma\,s_b}\right),\qquad
  \mathcal{Q}(X;\mathcal{B})_i=\gamma\,s_b\,x_i,\qquad i\in b,
  \label{eq:nvfp4}
\end{equation}
with $x_i=0$ whenever $s_b=0$. We identify an NVFP4 tensor with the real tensor that it represents, so sums and products of hatted (quantized) tensors denote exact real arithmetic. Block-scaled Tensor Core instructions evaluate these products from the stored values and scales with FP32 accumulation, without materializing a wider tensor. $\mathcal{Q}_{1\times16}$ and $\mathcal{Q}_{16\times16}$ apply Equation~\ref{eq:nvfp4} with $1\times16$ and $16\times16$ partitions, respectively. The tensor scale $\gamma$ is a per-layer constant for each operand; for queries and the cache in both models, $\gamma=\eta\,X_{\max}/(M_{\mathrm{E2M1}}M_{\mathrm{E4M3}})$ is fixed offline from the largest magnitude $X_{\max}$ observed during calibration, with margin $\eta>1$.

\section{Arithmetic intensity of MLA decoding}
\label{app:intensity}

MLA's compact shared cache and wider BMMs after matrix absorption raise arithmetic intensity and make math throughput relevant alongside memory bandwidth.

\subsection{A shared cache supports wider BMMs}
Let $T$ denote the cached-token count, $n_h$ the number of query heads, and $n_g$ the number of GQA KV groups. With key and value widths $d_K$ and $d_V$, one GQA query performs $F$ FLOPs on $N$ distinct cache elements:
\begin{equation}
  F_{\mathrm{GQA}}=2n_hT(d_K+d_V),\qquad
  N_{\mathrm{GQA}}=n_gT(d_K+d_V).
  \label{eq:gqa-work-payload}
\end{equation}
For MLA, BMM1 uses $d_c+d_{\mathrm R}$ channels and BMM2 uses $d_c$:
\begin{equation}
  F_{\mathrm{MLA}}=2n_hT(2d_c+d_{\mathrm R}),\qquad
  N_{\mathrm{MLA}}=T(d_c+d_{\mathrm R}).
  \label{eq:mla-flops}
\end{equation}
The NoPE latent contributes to both products but occupies storage only once.

For DeepSeek's $n_h=128$, $d_c=512$, and $d_{\mathrm R}=64$~\citep{deepseekv3}, we compare a GQA reference with $n_g=8$ and $d_K=d_V=128$. At equal payload bytes $\beta$ per element, ideal intensity is $I^\star=F/(\beta N)$. Assuming one cold-cache read with reuse across heads and, for MLA, across BMM1 and BMM2 gives
\begin{equation}
  \frac{F_{\mathrm{MLA}}}{F_{\mathrm{GQA}}}=4.25,\qquad
  \frac{N_{\mathrm{GQA}}}{N_{\mathrm{MLA}}}\approx3.56,\qquad
  \frac{I^\star_{\mathrm{MLA}}}{I^\star_{\mathrm{GQA}}}=\frac{136}{9}\approx15.1.
  \label{eq:intensity-ratio}
\end{equation}

\subsection{MTP raises intensity through cache reuse}
Let $n_q$ query positions share a long prefix in one invocation, including the ordinary next-token position. Neglecting newly appended positions gives $F(n_q)\approx n_qF(1)$. With $\kappa(n_q)=D_{\mathrm{cache}}(n_q)/D_{\mathrm{cache}}(1)$ as cache traffic relative to a single query, excluding query-dependent transfers gives
\begin{equation}
  I(n_q)\approx\frac{n_q}{\kappa(n_q)}I(1).
  \label{eq:mtp-intensity}
\end{equation}
Full reuse gives $\kappa(n_q)=1$ and an approximately $n_q$-fold increase; MTP=3 evaluates four query positions.

\subsection{FP4 addresses both bandwidth and math throughput}
For matrix-throughput ceiling $\pi_{\mathrm{mat}}$, HBM bandwidth $\omega_{\mathrm{HBM}}$, and HBM traffic $D_{\mathrm{HBM}}$, roofline analysis gives the latency lower bound~\citep{williams2009roofline}
\begin{equation}
  t\geq\max\!\left(\frac{F}{\pi_{\mathrm{mat}}},\frac{D_{\mathrm{HBM}}}{\omega_{\mathrm{HBM}}}\right).
  \label{eq:mla-roofline}
\end{equation}
FP4 storage reduces the memory term, and native FP4 BMMs reduce the compute term. With MTP, FP8 MLA decoding is compute-bound on the evaluated GPU (Section~\ref{sec:eval-kernel}).

\section{Block-local softmax-score scaling}
\label{app:probability-scaling}

Section~\ref{sec:implementation} computes the E2M1 score values $x_i$ and the block scales $s_b$ of Equation~\ref{eq:p-fma-form}. Once $m$ is available, the kernel encodes the block scale with the tensor scale $\gamma_P=1/(M_{\mathrm{E2M1}}M_{\mathrm{E4M3}})$ of Equation~\ref{eq:nvfp4}:
\begin{equation}
  s_b=\mathcal{Q}_{\mathrm{E4M3}}\!\left(M_{\mathrm{E4M3}}\,2^{a(m_b-m)}\right),
  \qquad
  \widehat{p}_i=\gamma_P\,s_b\,x_i,\quad i\in b,
  \label{eq:p-scale}
\end{equation}
where $\mathcal{Q}_{\mathrm{E4M3}}$ rounds to nearest with ties to even, as the native E4M3 conversion does. Because the values $x_i$ assume the unrounded scale, scale rounding multiplies all scores of a block by the common factor $s_b/\bigl(M_{\mathrm{E4M3}}\,2^{a(m_b-m)}\bigr)$. Rather than absorbing $1/\ell$ into every block scale, the kernel keeps this reciprocal off the BMM2 critical path and divides the BMM2 output by $\ell$ after BMM2, with $\ell$ accumulated in FP32 from the unquantized scores $\tilde{p}_i$.

\paragraph{Range and underflow of the softmax-score scale.}
Because $m_b\le m$, the argument of $\mathcal{Q}_{\mathrm{E4M3}}$ in Equation~\ref{eq:p-scale} lies in $(0,M_{\mathrm{E4M3}}]$: $\gamma_P$ maps the row maximum to the largest finite E4M3 value, and no block scale saturates. In the normal E4M3 range, $\mathcal{Q}_{\mathrm{E4M3}}$ changes a block's scores by a common factor within $1\pm2^{-4}$. A block falls into the subnormal range, where the scale spacing is $2^{-9}$, when $2^{a(m_b-m)}<2^{-6}/M_{\mathrm{E4M3}}\approx3.5\times10^{-5}$. Round-to-nearest flushes the scale to zero when
\begin{equation}
  2^{a(m_b-m)}\le\tau=\frac{2^{-10}}{M_{\mathrm{E4M3}}}\approx2.2\times10^{-6},
  \label{eq:p-underflow}
\end{equation}
that is, only when every score in the block is at most $\tau$ times the row maximum. Let $\mathcal{F}$ be the set of tokens in flushed blocks. Because $\tilde{p}_i\le\tau$ for $i\in\mathcal{F}$ and $\ell\ge1$, flushing removes at most
\begin{equation}
  \frac{1}{\ell}\sum_{i\in\mathcal{F}}\tilde{p}_i\le\lvert\mathcal{F}\rvert\,\tau
\end{equation}
of the normalized softmax mass. Within every retained block, E2M1 rounding already maps arguments of $\mathcal{Q}_{\mathrm{E2M1}}$ at or below $0.25$ to zero and thus removes any score at or below $1/24$ of its block maximum.

\end{document}